\documentclass[twocolumn,aps,prb]{revtex4}
 
\usepackage{graphicx}
\usepackage{epsfig}

\begin{document}
\baselineskip11pt
\parindent0pt 

\title{Universal vortex formation in rotating traps with bosons and fermions
} 

\bigskip
 
\author{M. Toreblad, M. Borgh, M. Koskinen$^*$, M. Manninen$^*$ and  \\
S.M. Reimann}

\bigskip

\address{Mathematical Physics, Lund Institute of Technology, SE-22100 Lund, Sweden}

\address{$^*$NanoScience Center, Department of Physics, FIN-40351 University of Jyv\"askyl\"a, Finland}
 
\begin{abstract}
When a system consisting of many interacting particles is set rotating,
it may form vortices. This is familiar to us from 
every-day life: you can observe vortices while stirring your coffee
or watching a hurricane.  
In the world of quantum mechanics, famous examples of vortices are 
superconducting films\cite{tinkham1996} and rotating 
bosonic $^4$He or fermionic $^3$He liquids\cite{yarmchuk1979,ruutu1996}. 
Vortices are also observed in rotating Bose-Einstein condensates
in atomic traps~\cite{matthews,madison2000,abo} and are
predicted to exist\cite{rodriguez2001} for paired fermionic 
atoms\cite{ferm1,ferm2}.
Here we show that the rotation of trapped particles with a repulsive 
interaction leads to a similar vortex formation, regardless of whether the
particles are bosons or (unpaired) fermions.~The exact, quantum
mechanical 
many-particle wave function provides
evidence that in fact, the mechanism of this vortex formation is 
the same for boson and fermion systems. 
\end{abstract}

\maketitle

Let us now consider a number of identical particles 
with repulsive interparticle interactions confined in a harmonic
trap under rotation. These particles could
be electrons in a quantum dot\cite{review}, positive or negative ions,
or neutral atoms in boson or fermion condensates\cite{chris}.
Though simple to describe, this 
quantum mechanical many-body problem is extremely complex and 
in general not solvable exactly. 
Consequently, in rotating systems the formation of vortices 
and their mutual interaction
is usually described using a mean field approximation. 
In superconductors this is the Ginzburg-Landau
method\cite{tinkham1996}. For Bose-Einstein condensates, 
one often applies the Gross-Pitaevskii
equation\cite{stringari,chris}. 
In this way, Butts and Rokhsar\cite{butts} found successive transitions 
between stable patterns of singly-quantised vortices, as the angular
momentum was increased. A single vortex appears when the angular
momentum $L$  is equal to the number of particles $N$, two vortices appear
at $L\sim 1.75N$ and three vortices at $L\sim 2.1N$
(see Refs.~\cite{butts,ben,bertsch1999,george}). 
For quantum dots in strong magnetic fields, the occurrence of vortices 
was very recently discussed by Saarikoski {\it et al.}~\cite{saarikoski2004}.

Based on the rigorous solution of the many-particle Hamiltonian, 
we show that striking similarities between the boson and
fermion systems exist: the vortex formation is indeed universal 
for both kinds of particles, and the many-particle configurations
generating these vortices are the same. 
For a small number of particles, the many-body Hamilton
operator can be diagonalised numerically. 
We use a single particle basis of Gaussian functions to span the
Hilbert space. These Gaussians are eigenstates of 
the trap for radial quantum number $n=0$ and different single-particle
angular momenta. These states 
dominate for large {\it total} angular momenta $L$.
We only consider one spin state, i.e.\ bosons with 
zero spin or spin-polarised fermions.
Numerical feasibility limits calculations to small particle numbers $N$.
However, the advantages of our approach as compared to
mean field methods are that our solutions (i)~are exact
(up to numerical accuracy),
(ii)~maintain the circular symmetry and thus have a good angular
momentum, and 
(iii)~allow the direct, quantitative comparison between boson and 
fermion states and thus serve to uncover the origin of the vortices in 
small systems.

The many-particle energies for rotating clouds of bosons or fermions 
are compared in Figs.~1 (bosons) and 2 (fermions). The low-lying states
are shown as a function of total angular momentum $L$. 
Following the tradition in nuclear physics,  the line connecting the 
lowest states at fixed $L$ is called\cite{ben}
the ``yrast'' line. 
(The word ``yrast'' originates from  Swedish 
language  and means ``the most dizzy''.)
It is marked with a red line in Figs.~1 and~2. 
For bosons the two spectra appear almost identical, although one of them
is calculated with a short-range contact 
interaction and the other with a long-range Coulomb interaction\cite{vorov2003}.

The comparison between the lowest energy states of the spectra 
for bosons and Coulomb-interacting fermions, as displayed in Figs.~1 and ~2, 
reveals striking similarities: 
The yrast line has the same kinks and vortices can be found in both
systems appearing at similar angular momenta, as we explain below. 

When studying the appearance of vortices in the boson or fermion
densities, we should remember that in contrast to mean field methods, 
for an exact calculation with good angular momentum 
the particle density has circular
symmetry and thus does not display the internal structure directly.
To find the vortices we therefore would need to study 
pair-correlation functions. In the fermion
case, however, this can be problematic due to the disturbance of 
the exchange hole. 
Alternatively, as done in the insets of Figs.~1 and~2, we
break the circular symmetry
with a small perturbation of the form $V_{\ell
}(r,\varphi)=V_0\cos(\ell \varphi)$
which has $\ell $ minima around the center.
The perturbation can only couple states which differ in angular
momentum by $\pm \ell $.  Since the lowest energy state for each
angular momentum is also the most important in the perturbation
expansion, we can estimate the effect
of the perturbation by the mixture of three yrast states
$\tilde \Psi(L)=\Psi_0(L)+\eta[\Psi_0(L-\ell )+\Psi_0(L+\ell )]$, 
where $L$ is the angular momentum and $\eta $ is the mixing
parameter, which is of the order 0.1 or smaller. 
If, for instance, the state in question has a two-vortex structure, then 
this will appear in $\tilde \Psi (L)$ as two distinct minima for $\ell
=2$ already at very small mixing ratio $\eta $.
\begin{figure}[h]
\centerline{\epsfxsize=3.4in\epsfbox{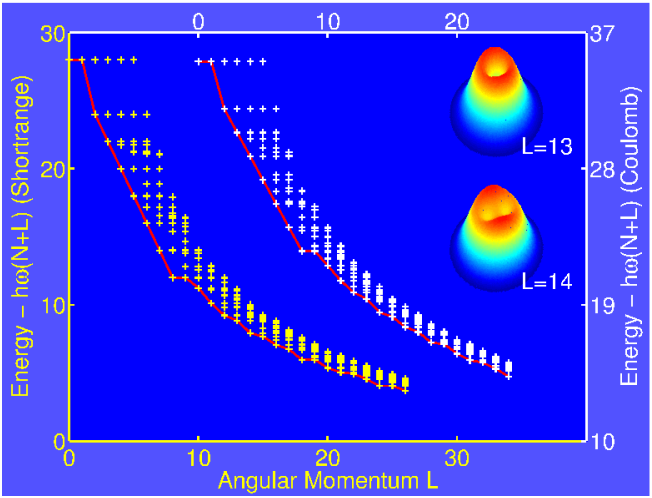}}
\vspace{0.5cm}   
\caption{Rotating bosons in a trap. The Figure shows the 
low-lying many-particle energies of $N=8$ 
bosons interacting by a contact interaction ({\it yellow, lower left}) or  eight charged
bosons interacting by the 
long-range Coulomb repulsion ({\it white, upper right}), as a function of angular
momentum $L$ (in units of $\hbar $). 
The read line, also called the ``Yrast line'', connects 
the lowest states at fixed $L$.  The insets show vortices
in the perturbative densities (as explained in the text), occurring 
above ratios $L/N=1$ for a single vortex and $L/N=1.75$ for two vortices.}
\end{figure}
\begin{figure}[h]
\centerline{\epsfxsize=3.3in\epsfbox{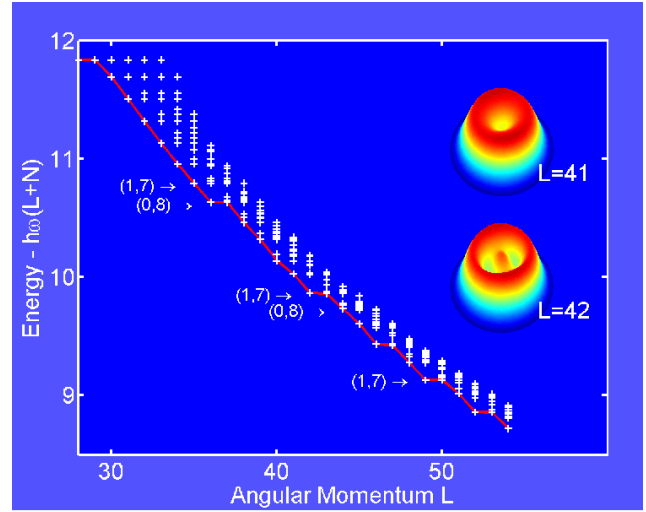}}
\vspace{0.5cm} 
\caption{Rotating, spin-polarised charged fermions (for example,
 electrons) in a trap. The Figure shows the low-lying many-particle
 energies for $N=8$. 
The similarity to the boson case is remarkable. 
If bosons have a vortex state at angular momentum $L_B$, the fermions 
can have the same vortex structure at angular momentum $L_B+N(N-1)/2$.
In the example shown here, for fermions 
a single vortex can be seen in the perturbative
densities above $(L-28)/N=1$. The inset shows a single vortex at $L=41$,
and two vortices at $L=42$, corresponding to 
$(L-28)/N=1.75$. }
\end{figure}
The insets to Figs.~1 and~2 are obtained in this way. In the boson
case, for N=8 the perturbative densities show a single vortex 
for $L=13$, corresponding to L/N=1.6, while a second vortex occurs at 
$L=14$, that is $L/N=1.75$.
For fermions, the angular momentum is shifted\cite{paredes} by $N(N-1)/2$
as explained below. In analogy to the boson case, for $N=8$ fermions
the single vortex still exists at $L=28+13$, while two vortices 
are found at $L=28+14$.

This universality in the vortex formation can be
understood by looking more in detail at 
the many-particle states of the rotating system.
In the case of non-interacting bosons
the many-particle ground state is
\begin{equation}
\Psi_B=e^{-\sum_k \vert z_k\vert^2},
\end{equation}
where the coordinates in two dimensions are expressed by 
complex numbers $z_j=x_j+iy_j$.
It turns out that in the case of spin-polarised Fermions the
corresponding ``condensate''  
is the so-called {\it maximum density droplet} (MDD)\cite{macdonald1993}
\begin{equation}
\Psi_{F}=\prod_{j<k}^N(z_j-z_k)e^{-\sum_l \vert z_l\vert^2}~,
\end{equation}
which is a
Slater determinant of the consecutive single particle states, filled
from $m=0$ to $m=N-1$ and thus has a non-zero angular momentum
$L_F=N(N-1)/2$. The state $\Psi _F $ corresponds to the Laughlin wave
function for the integer quantum Hall effect\cite{laughlin1983}.

For bosons with short range repulsive 
interaction, a single vortex can be formed by multiplying the boson
ground state by a symmetric polynomial\cite{wilkin,bertsch1999}
$P_{1V}=\prod_k^N (z_k-z_0)$,
where $z_0$ is the center of mass.
If one multiplies the MDD with the same polynomial\cite{manninen2001}, 
this gives a good approximation for the exact single vortex state for charged
fermions.
By noticing that for a system with many particles, the center of mass
can be put at the origin, $z_0=0$, we can make an Ansatz for the 
state with $n$ fixed vortices forming a ring around the origin:
\begin{eqnarray}
\Psi_{nV} &=&\prod_{j_1}^N (z_{j_1}-ae^{i\alpha_1})\times \cdots\times
\prod_{j_n}^N (z_{j_n}-ae^{i\alpha_n}) \Psi_{B,F} \nonumber\\
&=& \prod_j^N (z_j^n-a^n)\Psi_{B,F}~,
\end{eqnarray}
where $\Psi_{B,F}$ is either the boson condensate or the fermion MDD, and
the vortex centers are localised on a ring of radius $a$
($\alpha_k=k\cdot {2\pi \over n}$).
This state does not have a good angular momentum, but such a state can be 
projected out by collecting only
terms corresponding to a specific power of the constant $a$,
\begin{equation}
\Psi_{nV}=a^{n(N-K)}{\cal{S}}\left(\prod_k^K z_k^n\right)\Psi_{B,F},
\end{equation}
where $\cal{S}$ means symmetrisation.
Note that with $n=1$, $K=N$ and $a=0$ this state describes a single
vortex fixed at the origin.
Figure~3 shows schematically 
the single particle occupation of these ``vortex generating states'' for
bosons and fermions. In both cases the $n$ vortices are generated by 
exciting $K$ single particles by $n$ units of angular momentum.
\begin{figure}[h]
\centerline{\epsfxsize=2.8in\epsfbox{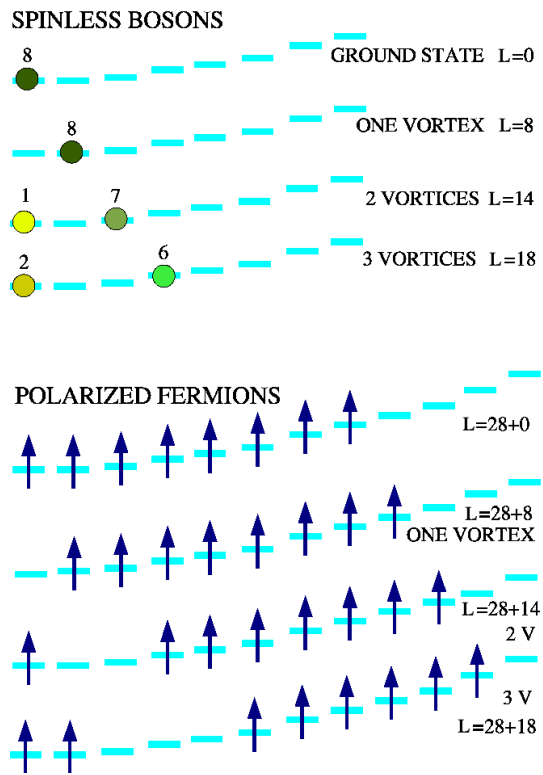}}
\caption{Vortex generating single-particle configurations for bosons
  and fermions.  (The particle number is chosen to be eight in this example). 
Exactly the same excitations of the ``ground states'' 
(condensate for bosons, and maximum density droplet for fermions) 
cause the vortices in both cases, with the
same increase in angular momentum.}
\end{figure}
The quantum states of the numerical exact solutions show that the dominating 
configurations in cases where we see 1, 2, or 3 vortices 
(independent of the number of particles) are indeed those shown
in Fig.~3, for both bosons and fermions. 

In the exact diagonalisation, by multiplying 
the exact boson wave function with the wave function  $\Psi_F$  
of the MDD, Eq.~(2), we can determine the overlap between the fermion and
boson states. It turns out to be even larger than the weights
of the most important configurations. For example,
the overlap between the two-vortex states shown in
Figs.~1 and~2 is 57\% while the weight of the
most important configuration (as shown in Fig.~3) 
in the boson case is only 15\%
and in the fermion case 47\%.

The vortices are born by the rotational motion and consequently
carry angular momentum. In the single particle picture the angular
momentum is associated with the phase of the complex wave function: 
Going around the angular momentum axis the phase changes by $2\pi$.
Similarly, the phase changes 
by $2\pi$ in going around a vortex center. In the many-particle
picture the phase of the wave function depends on the coordinates of
all the particles.
In this case, the phase change around the 
vortex cores can be visualised by fixing
the coordinates of $N-1$ particles and plotting the phase as a function
of the last coordinate~\cite{saarikoski2004}. 
This is done in Fig.~4 for the vortex generating configuration
for $N=8$. The state $\Psi_F$ has maximum {\it amplitude}, when the
electrons are located so that one electron is in the center and seven 
electrons form a ring around it.
To study the phase, we fix six of the electrons on the ring and one at
a slightly off-center position in the middle. The resulting phase is shown
in Fig.~4a. One can clearly see 
that each electron carries a vortex with it, as known
from the theory of the integer quantum Hall effect\cite{prange}. 
When the wave function is multiplied with the polynomial generating 
the vortices, Eq. (4) with $n=2$, two additional vortices appear
(Fig.~4b). When the fermion state $\Psi_F$ is replaced with 
the corresponding boson state 
$\Psi_B$ only the two additional vortices are seen (Fig.~4c).
\begin{figure}[h]
\vspace{-5cm}
\centerline{\epsfxsize=3.8in\epsfbox{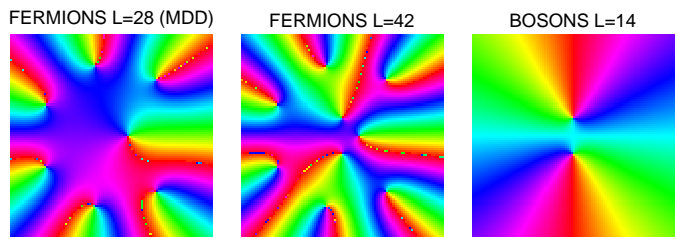}}
\vspace{-5cm}   
\caption{The phase of the many-particle wave function for eight
  particles. 
The phase is shown as a function of the coordinates of one particle
  with all other coordinates fixed. The colour scale is such that 
the jump from $-\pi $ to $\pi $ corresponds to a change from
  {\it orange} to {\it pink}. 
(a) shows the phase of the fermion state of the maximum 
density droplet  $\Psi_F$, Eq.~(2), 
with vortices localised at the fixed electrons,
and (b) shows the phase of the two-vortex state.
The two additional vortices appear in $\Psi _{2V}$, when the
state in (a) is multiplied by the vortex generating polynomial, see
  Eq.~(4), and 
(c) shows the same two vortices
for the boson state.}
\end{figure}

Finally, we will return to the fermion spectrum shown in Fig.~2.
The maximum amplitude of the MDD, i.e. the 
fermion ``condensate'', corresponds to the equilibrium particle
positions of a classical system with logarithmic repulsive
interactions\cite{laughlin1983}. In small systems 
a rigid rotation of this localised state gives some of the 
high angular momentum states\cite{manninen2001,maksym,jain}. 
For example, in the case of eight particles
there are two classical configurations: a single ring of eight
particles, which we label by $(0,8)$, 
and a ring of seven particles with one particle at the
center, $(1,7)$. 
The former allows rigid rotation at angular momenta  
$28+8=36$, $28+16=44$, etc., while the latter at $28+7=35$, 
$28+14=42$, etc. (as marked by arrows in Fig.~2). 
In the fermion systems, in some cases both localisation and 
vortex structure coincide. This is for example 
the case at $L=42$ for eight fermions, 
which in the pair correlation function 
shows two vortices as well as very weakly localised particles
arranged in a (1,7)-configuration.

To summarise,
we have shown with exact solutions of the many-particle systems that
the vortex formation in rotating traps of bosons and fermions 
have universal features: (i) They appear at certain angular momenta
determined by the number of particles and number of vortices,
(ii) the many-particle excitations generating the vortices are
the same for bosons and fermions, and (iii) the vortex formation does not
depend on the shape of the repulsive interaction between the particles.

\end{document}